# The optical microscopy with virtual image breaks a record: 50-nm resolution imaging is demonstrated

Zengbo Wang[1*], Wei Guo[1,2], Lin Li[1], Zhu Liu[2], Boris Luk'yanchuk[3], Zaichun Chen[4] and Minghui Hong[3,4]

[1] Laser Processing Research Centre, School of Mechanical, Aerospace, and Civil Engineering, University of Manchester, Manchester M60 1QD, UK
[2] Corrosion and Protection Centre, School of Materials, University of Manchester, The Mill, University of Manchester, Manchester M60 1QD, UK
[3] Data Storage Institute, DSI Building, 5 Engineering Drive 1, Singapore, 117608
[4] Department of Electrical and Computer Engineering, National University of Singapore, Singapore 117576
*E-mail: zengbo.wang@manchester.ac.uk

**We demonstrate a new 'microsphere nanoscope' that uses ordinary SiO2 microspheres as superlenses to create a *virtual image* of the object in near field. The magnified virtual image greatly overcomes the diffraction limit. We are able to resolve clearly 50-nm objects under a standard white light source in both transmission and reflection modes. The resolution $\lambda/12$ achieved for white light opens a new opportunity to image viruses, DNA and molecules in real time.**

Galileo Galilei invented an optical microscope in 1609. He developed an *occhiolino* or a compound microscope with a convex and a concave lens. Although Galileo was probably not the first inventor of a microscope, after his invention the microscope became a commonly used and powerful scientific tool. There is no doubt that this invention is among the most important scientific achievements in the history of mankind. A quarter of a millennium later, Ernst Abbe established the formula for the resolution limit of the microscope: the minimum distance for two structural elements to be imaged as two objects instead of one is given by $d = \lambda / NA$, where $\lambda$ is the wavelength of light and $NA$ is the numerical aperture of the objective lens. The physical origin for this limited resolution is related to diffraction and the loss of evanescent waves in the far field. These evanescent waves carry high spatial frequency sub-wavelength information of an object and decay exponentially with distance. For about one hundred years, this Abbe criterion was considered as the fundamental limit of microscope resolution.

The next important step in optical microscopy was made with the help of near-field optics. In this case resolution of the image is limited by the size of the detector aperture and not by the wavelength of the illuminating light. The basic idea of near field optics can be easily understood from the Heisenberg's uncertainty principle: having fast decay of evanescent wave in z-direction one can greatly enhance resolution in x-y directions. Note that the famous Abbe sine condition also can be understood on the basis of uncertainty principle. The near-field scanning optical microscope (NSOM/SNOM) enables a resolution of a few ten nanometers[1]. However, it has many limitations, related to a very low working distance and extremely shallow depth of field. Also the transmission efficiency of a small aperture is very low. It also takes a long time for scanning over a large sample area for a high resolution imaging.

The diffraction limit can be also overcome by some other techniques, e.g. with surface-plasmon superlenses[2], nanoscale solid-immersion-lens[3], and molecular fluorescence microscopy[4,5]. The new concept of super resolution came from John Pendry, who suggested the basic idea of the "superlens"[6], related to the enhancement of evanescent waves within a slab of artificial material with a negative refractive index. Through the resonant coupling of evanescence waves to surface plasmon polaritons (SPP) in silver, objects as small as 60 nm were successfully recorded in the UV spectrum ($\lambda/6$ near-field resolution at $\lambda$ = 365 nm)[7]. Soon after, a SiC superlens, working at the mid-infrared frequency range ($\lambda$ = 11 μm) was demonstrated to resolve 540-nm holes[8]. However, these superlenses are 'near-sighted', since the images can only be picked up in the near field. In order to project near-field images into the far field, far-field superlenses (FSL) was shortly proposed and demonstrated[9,10]. The FSL used a silver slab to enhance the evanescent waves and an attached line grating to convert the evanescent waves into the propagating waves into the far field. However, such FSL did not magnify objects. In order to produce a magnifying superlens Smolyaninov et al[11] suggested the use of two-dimensional SPP confined by a concentric polymer grating placed on a gold surface. It generates 3x magnification, and a resolution of 70 nm at 495-nm wavelength ($\lambda$/7 far-field resolution).

Hyperlens is another type of magnifying superlens. The hyperlens used an anisotropic medium with a hyperbolic dispersion that generates a magnification effect through the cylindrical curved multilayer stacks[10]. The hyperlens worked at 365-nm wavelength and achieved a resolution of 130 nm ($\lambda$/3 far-field resolution) with <3x magnification[12]. Due to the SPP energy loss and sophisticated nanofabrication process, the resolutions of existing SPP-superlens and hyperlens are limited within $\lambda$/7 to $\lambda$/3. Moreover, both magnification and resolution of these lenses are orientation-dependant, which implies that the final images are not isotropic within the imaging plane. The other practical limit is that the SPP-superlenses must be excited with a specific laser source and configuration (wavelength, polarization, incident angle). It is hard to achieve SPP-superlens function with a standard white light source.

Our idea of super resolution microscopy uses the near-field magnification in the virtual image by a convention optical microscope. It is well-known that transparent micro/nano-spherical particle can be used for near-field focusing, see in Fig. 1a. In Ref. 13 the ability to focus light by such spherical particle into focal size $\lambda/8$ was demonstrated experimentally. Since such a lens can focus light into sub-diffractional size in the near field, one can expect the reciprocal effect: near-field magnification of sub-diffractional object into a size of the order of $\lambda$. Although the idea of the object magnification with spherical particles was discussed previously[14] it was not clear up to which extend one can produce magnification and what is the mechanism of the image formation. When the size of the particle is much bigger than $\lambda$, the magnification effect can be illustrated with the help of geometrical optics, see in Fig. 1b. The virtual image of a small object for this case is magnified by a factor of $n/(2-n)$ for $1 < n < 2$. Naturally for a sub-diffractional object in the near field, the geometrical optics is not applicable and one should solve Maxwell equations. However, as a reciprocal effect to the Mie theory, one can expect that an object with size in the order of at least $\lambda/8$ can be converted into image with a size $\lambda$, which can be seen by a conventional optical microscope. To the best of our knowledge, the idea of *near field magnification of the virtual image* has not been discussed previously. All the methods discussed above were operated with *real image*.

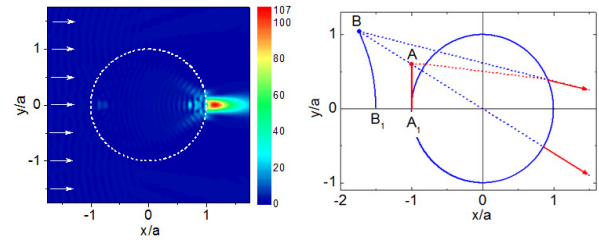

**Figure 1 | Light focusing and virtual image formation by a small dielectric sphere. a,** Field distribution for the near-field focusing of a 4.74 μm $SiO_2$ microsphere with a refractive index $n$ = 1.46 under a 600 nm light illumination calculated by the Mie theory. **b,** Formation of a virtual image at the limit of geometrical optics.

Here we need to mention that Plano-Spherical-Convex (PSC) shaped nanolenses were demonstrated recently for subwavelength imaging. It resolves 220-nm-line objects at 475-nm imaging wavelength ($\lambda$/2.2 far-field resolution, 2x magnification)[3]. These nanolenses, however, can be regarded as miniaturized solid immersion lenses (SILs)[15,16] because of the close geometrical profile and imaging resolution between them. In principle, our idea about the creation of a magnified virtual image can also be achieved with some special design of PSC in order to create much higher resolution, but to the best of our knowledge it has not been discussed previously.

Although it was not evident beforehand that one can produce a near field magnification of the virtual image, it was clear however that such an image can be formed by an evanescent wave only. From our experiments discussed below, we found that we can observe this magnified image by conventional optical microscope if we put the microscope image capture plane inside the particle at a distance from the top of the sphere. Thus, it was important to prove that the evanescent wave transfers high spatial frequency subwavelength information up to the position of the image capture plane. It can be verified from the exact solution of the Maxwell equations. We performed such calculations by CST Microwave Studio software for the particle on the surface of metallic periodical relief (perfect electric conductor layer with groove depth of 20 nm), see Fig. 2a. Calculations reveal clear formation of the periodical electric field distribution at an image plane, see in Fig. 2b. Once again we should emphasize that it is a distribution of "true light" but not the virtual image, e.g. we cannot see the object magnification. The virtual near-field image is formed on the other "virtual image plane" as it is discussed in the supplementary.

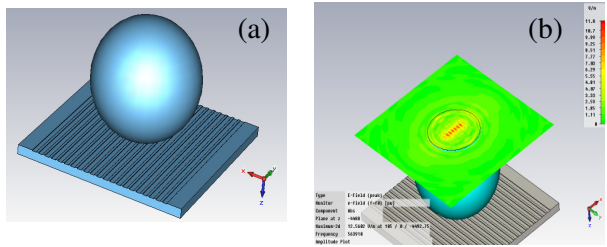

**Figure 2 | Microsphere transfers evanescent wave with high spatial frequency subwavelength information. a,** computer model of a 4.76-μm particle sitting on metallic periodical relief. Incident light is coming from the top and is *x*-polarized. **b,** electric field distribution at a plane across the top half of the particle where a grating image with same period of object appears.

In most microscopes, the eyepiece is a compound lens, with one component lens near the front and one near the back of the eyepiece tube. This forms an air-separated couplet. In typical design, the virtual image comes to a focus between the two lenses of the eyepiece, the first lens bringing the real image to a focus and the second lens enabling the eye to focus on the virtual image. However, in our case the situation is different, the virtual image plane is situated behind the object. That is similar to the situation when the virtual image is created behind the mirror on the distance to which the true light does not penetrate.

Figure 3a presents the schematic of our experiment in transmission mode. The microspheres are placed on top of the object surface by self-assembly[17]. A halogen lamp with a peak wavelength of 600 nm is used as the white light illumination source. The microspheres collect the underlying object information, magnify and project image to the far field. The final image was picked up by a conventional 80x objective lens (numerical aperture $NA$ = 0.9). In our initial experiments, gratings consisting of 360-nm-wide lines spaced 130 nm apart. It was imaged using 4.74-μm-diameter microspheres. We found that the images capture plane is positioned somewhere inside the top of the sphere (Fig. 3). This also reduces the light interference effect caused by neighbouring particles. As seen from Fig. 3b, only lines with particles on top of them have been resolved. The lines without particles on top are mixing together and forming a bright spot which cannot be directly resolved by the optical microscope because of its diffraction limit ($\lambda$/NA = 667 nm for $\lambda$ = 600 nm). The magnified image in Fig. 3b corresponds to a 4.15x magnification factor.

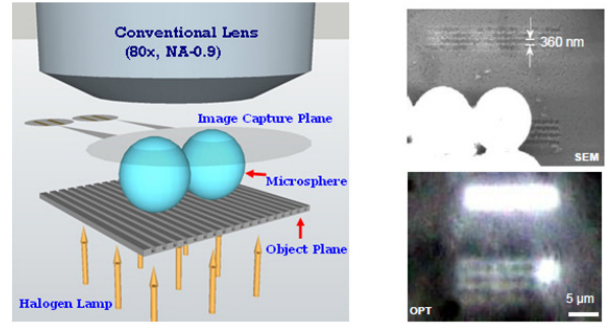

**Figure 3 | Experimental configuration of transmission mode microsphere nanoscope and its demonstration for subwavelength imaging. a,** Schematic of the transmission mode microsphere nanoscope system. Light coming from the bottom incidents on nanostructures and excites an evanescent wave. The spheres collect the evanescent wave information, which is transmitted through the sphere and captured by the conventional lens. **b,** Two photos presented images of 360-nm-wide lines spaced 130 nm apart from the SEM (upper photo) and optical image (bottom) which can be seen on the image capture plane of the microscope. The spacing of 130 nm is clearly resolved.

In Fig. 4 one can see a fishnet gold-coated anodic aluminium oxide (AAO) sample imaged with 4.74-μm-diameter microspheres. The pores are 50 nm in diameter and spaced 50 nm apart. The microsphere nanoscope clearly resolves these tiny 50-nm pores and magnifies them to 410 nm. It corresponds to super-resolution of $\lambda/12$, which is six times smaller than the optical diffraction limit.

It is important to note that the magnification factor in Fig.4 is almost two times of that in previous grating samples in Fig. 3b. This implies the performance of microsphere superlens is greatly affected by the near-field interaction of sphere and the substrate. In our experiment, we have confirmed that the gold coating layer not only enhanced the resolving power but also increased the magnification factor of microsphere superlens. It indicates that the plasmon resonance associated with gold material has played an important role. Here we should pay attention to the peculiarities of the Poynting vector distribution in the near field. Near the surface of plasmonic material these field lines have significant bending[18] in contrast to the straight light tracing at the limit of geometrical optics. These "bending rays" can greatly influence magnification of the virtual image.

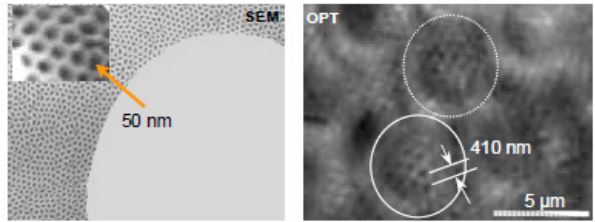

**Figure 4 | Microsphere nanoscope achieves 50-nm resolution under white light. a,** Microsphere on the top of a gold-coated fishnet anodic aluminium oxide (AAO) sample. The insertion shows the magnified AAO structure from SEM. **b,** Optical image of AAO structure with the microsphere nanoscope (radius of microspheres $a$ = 2.37 μm, borders of the spheres are shown by white lines). The nanoscope clearly resolves the pores which are 50 nm in diameter and spaced 50 nm apart. The size of the image between the pores within the image plane is 410 nm. It corresponds to a magnification factor of $M \approx 8.2$.

Since self-assembled particles are easy to spread over a large surface area, the images produced by each particle can be stitched together to form a large image. This is the case seen in Fig. 3b and Fig. 4 where a hexagonal array of particles functions as an array of superlens covering a large area.

Our previous calculations also reveal the fact that the Poynting vector of the radiation reflected by the surface transfers through the particle[19], which permits the formation of form an image in the reflection mode as well. Thus, we check that microsphere superlens also works for opaque materials in the reflection mode. Figure 5a demonstrates a Blu-ray DVD disk (200-nm-wide lines separated 100 nm apart) imaged with 4.74-μm-diameter microspheres in the reflection mode using the halogen light illumination. The sub-diffraction-limited lines are clearly observed. Figure 5b shows another example of reflection mode imaging of a nano-star structure made on a GeSbTe DVD disk. The complex shape of the star, including the 90-nm corners of the star, was clearly resolved by the microsphere superlens. It further proves the super-resolution capability of this microsphere nanoscope to overcome the optical diffraction limit.

Once again, we have to emphasize that the superior imaging performance demonstrated by the microsphere superlens is closely related to the remarkably short focal length and high near-field field enhancement effect arising from near-field focusing of light by microspheres. When the objects were illuminated, the evanescent waves containing sub-diffraction-limited object information were brought to a near-field interaction with the microspheres sitting on top of them, leading to the resonant coupling of evanescent modes to spherical cavity resonances (SCR) modes. It is also important that we resolve the magnified virtual image. The real image, produced by evanescent wave just increases the resolution but has no magnification as one can see in Fig. 2b. Another important conclusion of experimental work is that to reveal the virtual image at 50 nm we need the thin coated metallic layer. We were not able to see the magnified image without such layer. We used thin 20 nm gold films in contrast to superlenses where a weakly dissipating silver layer produces enhancement of evanescent wave[7] (note, that such superlenses increase resolution and do not produce magnification of the image).

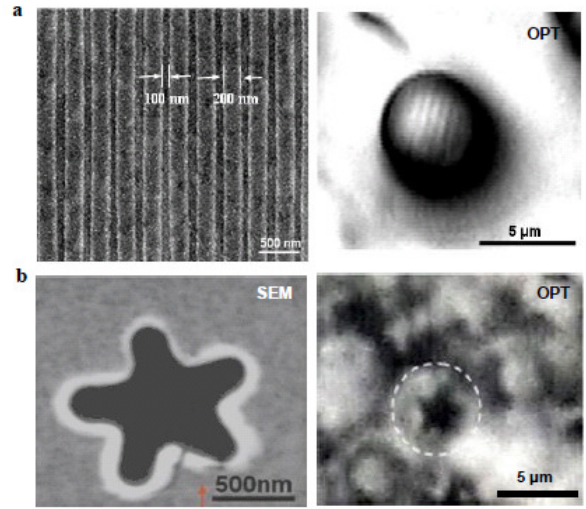

**Figure 5 | Reflection mode imaging. a,** Microsphere superlens reflection mode imaging of a commercial Blu-ray DVD disk. The 100-μm-thick transparent protection layer of the disk was peeled off before applying the microsphere (a = 2.37 μm). The sub-diffraction-limited lines are resolved by the microsphere superlens. **b,** Reflection mode imaging of a star structure made on GeSbTe DVD disk. The complex shape of the star including 90-nm corner was clearly imaged.

In summary, we have demonstrated that optically transparent microspheres are high performance optical superlenses for virtual image, which could resolve 50-nm objects under a white light source illumination. The microsphere superlens can be easily integrated with a standard optical microscope in either the transmission or the reflection mode to form a super-resolution optical nanoscope with $\lambda/12$ resolution. The microsphere nanoscope is robust, economical and it is also easy to accommodate different kinds of samples. Also, we can see a clear strategy for further enhancement of magnification factor with variation of refractive index and particle size. According to theoretical estimations, with refractive index close to $n <\approx 2$, it is possible to enhance the resolution of optical microscope with virtual image up to 10 nm.

**METHODS**

The gratings used in Fig. 3b were fabricated using a focused ion beam machine (Quanta 200 3D, FEI). The sample consists of 30-nm-thick chrome (Cr) film coated on fused silica substrates. The AAO used in Fig. 4 was fabricated by two steps anodizing in Oxalic Acid (0.3mol/l) under a constant voltage of 40V. A porous 20-nm-thick metallic film (gold) was then formed by using 300-μm-thick AAO as the template. The diluted $SiO_2$ sphere suspension was applied onto the substrate surface by drop coating. The colloidal silica spheres (Bangs Laboratories) form an ordered monolayer via self-assembly. An ordinary Olympus microscope was used to focus into the microsphere on the top half and images were collected and reported. The geometrical optical ray tracing analysis was done using Mathematica 7.0 software.

**Acknowledgements**
We thank Ms. Jie Guan for assisting in AAO sample fabrication. This work is supported by ASTAR/SERC Grant 0921540099.

**Author Contributions**
Z.W prepared the initial manuscript and conducted theoretical analysis and part of the imaging experiments. W.G conducted most the imaging experiments. L.L initialized the idea of nano-imaging using small spheres and L.L supervised the whole project. Z.L supervised nanoimaging experiments. B.L improved the manuscript presentation and imaging mechanism. Z.C and M. Hong contributed to the samples design and fabrication. All authors contributed to the analysis of this manuscript.

**Financial Interest Statement**
The authors have no competing interests as defined by Nature Publishing Group, or other interests that might be perceived to influence the results and discussion reported in this paper.

**References**

1. Hecht B., Sick B., Wild U.P., Deckert V., Zenobi R., Martin O.J.F. & D.W. Dieter, Scanning near-field optical microscopy with aperture probes: Fundamentals and applications, J. Chem. Phys. **112**, 7761-7774 (2000).

2. Zhang, X. & Liu, Z. W. Superlenses to overcome the diffraction limit. Nat Mater **7**, 435-441 (2008).

3. Lee, J. Y. et al. Near-field focusing and magnification through self-assembled nanoscale spherical lenses. Nature **460**, 498-501 (2009).

4. Hell, S. W., Schmidt, R. & Egner, A. Diffraction-unlimited three-dimensional optical nanoscopy with opposing lenses. Nat Photonics **3**, 381-387 (2009).

5. Hell, S. W. Far-field optical nanoscopy. Science **316**, 1153-1158 (2007).

6. Pendry, J. B. Negative refraction makes a perfect lens. Phys Rev Lett **85**, 3966-3969 (2000).

7. Fang, N., Lee, H., Sun, C. & Zhang, X. Sub-diffraction-limited optical imaging with a silver superlens. Science **308**, 534-537 (2005).

8. Taubner, T., Korobkin, D., Urzhumov, Y., Shvets, G. & Hillenbrand, R. Near-field microscopy through a SiC superlens. Science **313**, 1595-1595 (2006).

9. Liu, Z. W. et al. Far-field optical superlens. Nano Lett **7**, 403-408 (2007).

10. Jacob, Z., Alekseyev, L. V. & Narimanov, E. Optical hyperlens: Far-field imaging beyond the diffraction limit. Opt Express **14**, 8247-8256 (2006).

11. Smolyaninov, I. I., Hung, Y. J. & Davis, C. C. Magnifying superlens in the visible frequency range. Science **315**, 1699-1701 (2007).

12. Liu, Z. W., Lee, H., Xiong, Y., Sun, C. & Zhang, X. Far-field optical hyperlens magnifying sub-diffraction-limited objects. Science **315**, 1686-686 (2007).

13. Huang, S. M. et al.. Pulsed laser-assisted surface structuring with optical near-field enhanced effects, J. Appl. Phys. **92**, 2495-2500 (2002).

14. Yakovlev, V. V., Luk'yanchuk, B. Multiplexed nanoscopic imaging. Laser Physics **14**, 1065-1071 (2004)

15. Wu, Q., Feke, G. D., Grober, R.D. & Ghislain, L.P. Realization of numerical aperture 2.0 using a gallium phosphide solid immersion lens. Appl Phys Lett **75**, 4064-4066 (1999).

16. Zhang, J., See, C. W. & Somekh, M. G. Imaging performance of widefield solid immersion lens microscopy. Appl Optics **46**, 4202-4208 (2007).

17. Whitesides, G. M. & Grzybowski, B. Self-assembly at all scales. Science **295**, 2418-2421 (2002).

18. Wang, Z. B. et al. Energy flows around a small particle investigated by classical Mie theory, Phys. Rev. B. **70**, 032427 (2004).

19. Luk`yanchuk B. S., Wang Z. B., Song W. D. & Hong M. H., Particle on surface: 3D -effects in dry laser cleaning, Appl. Phys. A **79**, 747–751 (2004).

## Supplementary Method:

It is easy to see the magnified virtual image of the object with size much larger than the wavelength $\lambda$, e.g. with the help of spherical aquarium with water or with glass sphere. At this case magnification factor can be easily find with the help of geometrical optics. Let us consider the sphere with radius $a$ and refractive index $n$ and the point source $A$ with coordinates $x_A = -a$, $y_A = y_0$.

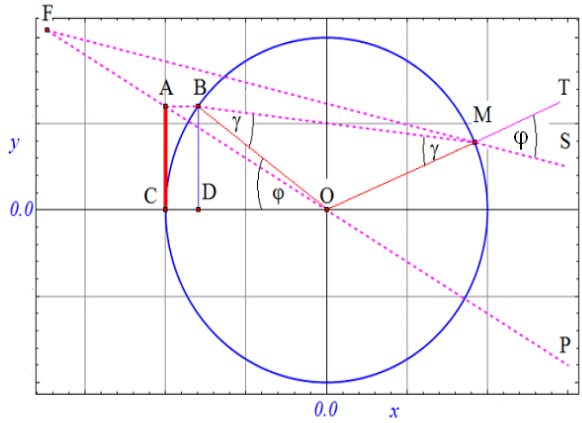

Fig. S1 | Image construction with dielectric sphere.

Ray AOP which go through the center of the particle have no deviation, thus the image of point $A$ is laying somewhere on the prolongation of this line. Ray AB which is parallel to axis $x$ enters into the particle at the point $B$ and has refraction. Then refracted ray moves up to point $M$, where it has the second refraction. After second refraction it goes along MS path. Image of the point $A$ is point $F$ where prolonged lines MS and AOP are crossed, see in Fig. S1. Using Snell's law $Sin\varphi = n\, Sin\gamma$ one can find coordinates of the M point: $x_M = a\, Cos(2\gamma - \varphi)$, $y_M = a\, Sin(2\gamma - \varphi)$, where $\varphi = ArcSin(y_0/a)$. After the second refraction the MS ray goes along the line presented by equation $y = y_M + (x_M - x)Tan[2(\varphi - \gamma)]$. Crossing of this line with line AOP, given by $y = -x\, y_0/a$ yields for coordinated of the image point F:

$$x_F = a\frac{x_M\, Tan[2(\varphi-\gamma)] + y_M}{a\, Tan[2(\varphi-\gamma)] - y_0},\quad y_F = -x_F\frac{y_0}{a}$$

For small values $y_0/a \ll 1$ we have the simplified formula

$$x_F \approx \frac{na}{2-n} - \frac{(n^3 - 2n + 1)y_0^2}{n(n-2)^2 a}.$$

One can see from (4) that at $n < 2$ increasing of $y_0$ leads to smaller $|x_F|$ value. As a result the image of strait object AC has bending, see in Fig. 1b.

Formation of the near-field virtual image is much more complicated. In Fig. S2 we present the Poynting vector lines for the microsphere with radius $a = 2.37$ μm and refractive index $n = 1.46$ illuminated by plane wave with $\lambda = 600$ nm.

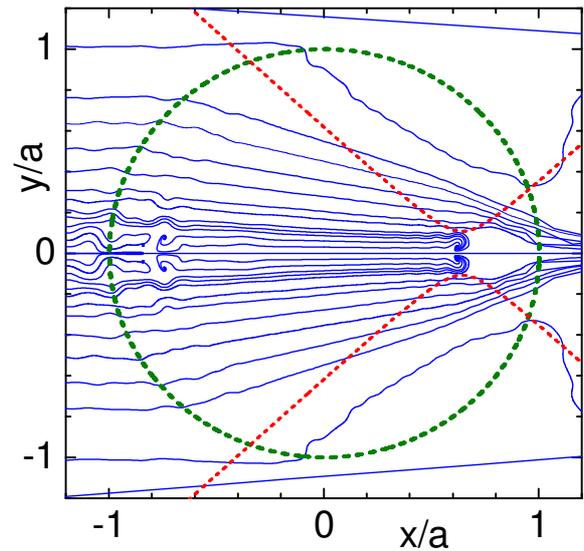

Fig. S2 | Poynting vector lines (blue) for the microsphere. Grin dash lines indicate the particle. Ref dot lines indicate the microscope caustics.

The time averaged Poynting vector is given by $\mathbf{S} = \frac{1}{2}\text{Re}[\mathbf{E}\times\mathbf{H}^*]$ and field lines are solution of differential equation $dx/S_x = dy/S_y$. We can see from the Figure that big objects with size $y/a > 0.2$ are transferred image more or less similar to geometrical optics. Small objects with $y/a < 0.2$ have different behaviour. They produce the energy vortex which is localized exactly at the caustics neck. "Very small objects" produce another vortex which do not reach the neck of the caustics at 2D plot. However from 3D Poynting vector lines we can see that they also reached focal points near $x/a \approx -0.75$. Thus, field to the vortexes within the caustic neck can arrive by two ways: either from the lines with $y/a <\approx 0.2$ or from the lines with much smaller size. In the limit of geometrical optics we should prolong stray rays. In the near field we should prolong lines of the Poynting vector. As a result we can see the virtual image identical to those which come from the real object with the size near $y/a \approx 0.2$. Characteristic magnification factor is given approximately by the ratio between two separatrixes which separates rays coming to the to two different vortexes. According to Fig. S2 it is about of factor 2.5. Now we can see a great difference between construction of the virtual image in the near field and in the case of geometrical optics. With geometrical optics one can see magnified virtual image though the sphere (i.e. we prolong rays which are out the sphere). In the case of near-field we form virtual image by prolonging Poynting vector lines inside the particle.